\documentclass[aps, prl, superscriptaddress, preprintnumbers, floatfix,twocolumn,10pt]{revtex4}
\usepackage{graphicx,amsfonts,amsmath,amssymb,amstext}
\usepackage{float,wrapfig}
\usepackage{subfigure, psfrag}
\usepackage{dsfont}
\usepackage{color}
\usepackage[colorlinks=true,urlcolor=blue,citecolor=blue,linkcolor=blue]{hyperref}
\usepackage{bm}

\newcommand{\be}{\begin{equation}}
\newcommand{\ee}{\end{equation}}
\newcommand{\ba}{\begin{eqnarray}}
\newcommand{\ea}{\end{eqnarray}}

\definecolor{red}{rgb}{0.7,0,0}
\definecolor{green}{rgb}{0,0.5,0}

\begin{document}

\title{Do we need  to use regularization on the thermal part in the NJL model? }
\date{\today}
\author{Kai Xue}
\thanks{co-first author}
\affiliation{School of Automotive and Traffic Engineering , Jiangsu University, Zhenjiang 212013 P.R. China}
\author{Xiaozhu Yu}
\thanks{co-first author}
\author{Xinyang Wang}
\thanks{wangxy@ujs.edu.cn, correspondence author}
\affiliation{Department of Physics, Jiangsu University, Zhenjiang 212013 P.R. China}
\begin{abstract}
The Nambu--Jona-Lasinio (NJL) model is one of the most useful tools to study non-perturbative strong interaction matter. Because it is a nonrenormalizable model, the choosing of regularization is a subtle issue. In this paper, we discuss one of the general things of the regularization in the NJL model, which is whether we need to  use the regularization on the thermal part by evaluating the quark chiral condensate and thermal properties in the two-flavor NJL model. The calculations in this work include three regularization schemes that contain both gauge covariant and invariant schemes. We found no matter which regularization scheme we choose, it is necessary to use the regularization on the thermal part when calculating the chiral condensate related physics quantities and do not use the regularization on the thermal part when calculating the grand potential related physical quantities.
\end{abstract}
\maketitle
\section{Introduction}
The understanding of QCD matter with finite temperature and chemical potential is one of the main topics in theoretical physics. In recent decades, the heavy-ion collision experiment especially RHIC and LHC create a ground experiment environment to study such matter. In the theoretical viewpoint, because the strong interaction coupling constant $\alpha_s$ is not small in most of the regimes, the perturbative method failed. Therefore, the non-perturbative methods are needed to understand the physics in the large $\alpha_s$ regime. One of the useful tools is the Nambu-Jona-Lasinio(NJL) model~\cite{Nambu:1961tp, Nambu:1961fr}. The NJL model has a long history back to 1961 (for reviews, see Refs.~\cite{Klevansky:1992qe, Buballa:2003qv, Hatsuda:1994pi}), it incorporates the chiral symmetry and its spontaneous breaking, the gluonic degrees of freedom are replaced by a local four-point interaction of color currents. The essential of this model is that it keeps the important symmetry in QCD, it is trivial to represent the symmetry breaking in QCD which is very useful for helping us to understand the properties of QCD matter. 

Because of the point interaction for quarks in the NJL model, the model is nonrenormalizable. Therefore, regularization should be used during the calculation. One of the fundamental problems is that when the temperature is finite, the contribution from the thermal part is convergent. It is problematic whether we need to use the regularization on this part the same as the divergent vacuum part. The different choice is not only to change the numerical values of the results  but also to change the physical behaviors. A good example is given in Ref.~\cite{Yu:2015hym}, where they were discussed by using/not using the regularization on the thermal part, the dependence of the critical temperature on the chiral chemical potential gave two opposite results. On the one hand, the thermal part does not need a regularization since it is convergent, to get the full contribution from the thermal part, we should not use the regularization on this part. On the other hand, regularization should be applied to all parts of the equation to get correct physics quantities under the same energy limit for physics consistency. Therefore, to answer this regularization of the thermal part problem, a systematic study is needed. 

In this paper, we calculate the quark chiral condensate/quark constituent mass and grand potential related thermodynamics by using the framework of  the two-flavor NJL model. We evaluate the physics quantities by using three different regularization schemes and introducing two different treatments on the thermal part for each scheme. We will investigate the simple and fundamental question: should we use the regularization on the thermal part when we are using the NJL model to calculate the physics quantities?

\section{model}
The Lagrangian density of the two-flavor NJL model is given by
\ba
\mathcal{L} = \bar{\psi}(i \gamma_{\mu}\partial^{\mu} - m) \psi + G_S\left[(\bar{\psi}\psi)^2+(\bar{\psi}i\gamma_5\mathbf{\tau}\psi)^2\right].
\ea
Here $m$ is the current quark mass, $\mathbf{\tau} = (\tau^{1}, \tau^{2}, \tau^{3})$ is the isospin Pauli matrices and  $G_{S}$ is the coupling constant with respect to the (pseudo)scalar channels.
By using mean-field (Hartree) approximation the Lagrangian density could be given by~\cite{Klevansky:1992qe},
\ba
\mathcal{L} = -\frac{\sigma^2}{4 G_S} + \bar{\psi}(i \gamma_{\mu}\partial^{\mu} - M) \psi,
\ea
where the dynamical quark mass $\sigma = -2G_S\left<\bar{\psi} \psi\right>$ and $\left<\bar{\psi} \psi\right>$ is the chiral condensate. Then the constituent quark mass is given by $M = m + \sigma$.  
The general grand potential can be written in vacuum part and thermal part,
\ba
\label{Omg-func}
\Omega&=&\Omega_{vac}+\Omega_{th},
\ea
where the vacuum and the thermal part are given by
\begin{subequations}
\ba
\Omega_{vac}=\frac{\sigma^{2}}{4G_{S}}-2N_f N_c \int \frac{d^3 p}{(2\pi)^3}E_p(\mathbf{p},M),
\ea
\ba
\Omega_{th}&=&-2N_f N_c \int \frac{d^3 p}{(2\pi)^3}T\nonumber\\
&\times&\left\{ \ln \left[1+ \exp\left(-\frac{E_p(\mathbf{p}, M)-\mu}{T}\right)\right]\right.\nonumber\\
&+&\left.\ln \left[1+ \exp\left(-\frac{E_p(\mathbf{p},M)+\mu}{T}\right)\right]\right\} .
\ea
\end{subequations}
Here,  the on shell energy of quark is given by $E_p(\mathbf{p},M) = \sqrt{\mathbf{p}^2 + M^2}$.  $N_f$ and $N_c$ are the number of flavours and colors, which are 2 and 3, respectively, in this work. The constituent mass/chiral condensate can be solved by the  corresponding gap equation $\frac{\partial \Omega}{\partial \sigma} = 0$ and  $\frac{\partial^2 \Omega}{\partial \sigma^2}>0$.
It is easy to see that the integral for the vacuum part  in Eq.~(\ref{Omg-func}) is divergent and the integral in the thermal part is convergent by integrating over the three momenta from 0 to infinity. 

In this study, we will use three different regularization schemes, (1) the three-momentum hard cutoff, (2) the three-momentum soft cutoff, and (3) Pauli-Villas regularization. One should be mentioned here, the first two schemes are gauge covariant and the third one is gauge invariant.

Then the vacuum part of grand potential with using regularizations could be written as 
\begin{subequations}
\label{vac}
\ba
\Omega^{H}_{vac}& =& \frac{\sigma^{2}}{4G_{S}}-N_f N_c \int_0^{\Lambda} \frac{d p}{\pi^2}p^2 E_p(p,M),
\ea
\ba
\Omega^{S}_{vac}& =& \frac{\sigma^{2}}{4G_{S}}-N_f N_c \int_0^{\infty} \frac{d p}{\pi^2}f_{\Lambda}^2p^2 E_p(p,M),\nonumber\\
\ea
\ba
\Omega^{PV}_{vac}& =& \frac{\sigma^{2}}{4G_{S}}-\sum_{a=0}^2 C_a \frac{N_f N_c}{\pi^2}\frac{M_a^4\ln M_a}{8}.
\ea
\end{subequations}

And the thermal part with using regularizations is given by
\begin{subequations}
\label{with}
\ba
\Omega^{H}_{th}& =& -N_f N_c \int_0^{\Lambda} \frac{d p}{\pi^2}p^2 T\nonumber\\
&\times&\left\{ \ln \left(1+ \exp\left(-\frac{E_p(p,M)-\mu}{T}\right)\right)\right.\nonumber\\
&+&\left. \ln \left(1+ \exp\left(-\frac{E_p(p,M)+\mu}{T}\right)\right)\right\},
\ea
\ba
\Omega^{S}_{th}& =& -N_f N_c \int_0^{\infty} \frac{d p}{\pi^2}p^2 f_{\Lambda}^2T\nonumber\\
&\times&\left\{ \ln \left(1+ \exp\left(-\frac{E_p(p,M)-\mu}{T}\right)\right)\right.\nonumber\\
&+&\left.\ln \left(1+ \exp\left(-\frac{E_p(p,M)+\mu}{T}\right)\right)\right\},
\ea
\ba
\Omega^{PV}_{th}& =& -\sum_{a=0}^2 C_a \frac{N_f N_c}{\pi^2} \int_0^{\infty} \frac{d p}{\pi^2}p^2 T\nonumber\\
&\times&\left[ \ln \left(1+ \exp\left(-\frac{E_p(p,M_a)-\mu}{T}\right)\right)\right.\nonumber\\
&+&\left.\ln \left(1+ \exp\left(-\frac{E_p(p,M_a)+\mu}{T}\right)\right)\right].
\ea
\end{subequations}
Where, the upper index H, S, PV represent hard cutoff, soft cutoff and Pauli-Villas, respectively. The overall grand potential with different regularizations and with/without applying the regularizations on the thermal part can be written by 
\ba
\Omega^{H}&=& \Omega^{H}_{vac} + \Omega^{H}_{th},~~\Omega^{H'} = \Omega^{H}_{vac} + \Omega_{th},\nonumber\\
\Omega^{S} &=& \Omega^{S}_{vac} + \Omega^{S}_{th},~~\Omega^{S'} = \Omega^{S}_{vac} + \Omega_{th},\nonumber\\
\Omega^{PV} &=& \Omega^{PV}_{vac} + \Omega^{PV}_{th},~~\Omega^{PV'} = \Omega^{PV}_{vac} + \Omega_{th},
\ea
where the primes represent regularization-free on thermal part. The general parameters are given in Table.~\ref{table:table1}. In addition, the soft cutoff weight function is chosen as 
\ba
f_{\Lambda}(p) =\sqrt{\frac{\Lambda^{2N}}{\Lambda^{2N}+|\mathbf{p}|^{2N}}},
\ea
here we use $N = 5$. For Pauli-Villas regularization, we set $C_{a} = (1,1,-2)$, $\alpha_a = (0, 2, 1)$ and $M_{a}^2 = M^2 + \alpha_{a} \Lambda^2$ for $a = (0, 1, 2)$, respectively. 
\begin{table}
\begin{tabular}{|c|c|c|c|}
\hline
~Regularization Scheme~ &~ $\Lambda$(MeV) ~& ~m(MeV)~& ~ G$_S$ ~ \\
\hline
three-momentum hard cutoff&653&6& $2.14/\Lambda^2$  \\
\hline
three-momentum soft cutoff&626.76&6&  $2.02/\Lambda^2$   \\
\hline
Pauli-Villas regularization&859& 6&  $2.84/\Lambda^2$    \\
\hline
\end{tabular}
\caption{The parameters of three different regularization schemes~\cite{Klevansky:1992qe, Fukushima:2010fe}.}
\label{table:table1}
\end{table}

Also, we want to introduce the thermodynamics quantities which depend on the grand potential directly in this work.The pressure  is just equal to the opposite sign of the grand potential, i.e., $P = -\Omega$. By introducing the normalized grand potential we will normalize the pressure at zero temperature and chemical potential equal to zero, then we have
\ba
P(\mu, T) = \Omega(0, 0)-\Omega(\mu, T),
\ea
where $\Omega(0, 0)$ is the grand potential in the vacuum. The energy density $\epsilon$ is given by
\ba
\epsilon = -T^2\left. \frac{\partial(\Omega/T)}{\partial T}\right|_{V}=-T\left.\frac{\partial \Omega}{\partial T}\right|_{V}+\Omega-\Omega(0, 0),
\ea
and the corresponding specific heat
\ba
C_{V} =\left.\frac{\partial \epsilon}{\partial T}\right|_{V} = -T\left.\frac{\partial^2 \Omega}{\partial T^2}\right|_{V}.
\ea
The square of velocity of sound at constant entropy $S$ is given by
\ba
v_s^2 = \left.\frac{\partial P}{\partial \epsilon}\right|_S = \left.\frac{\partial \Omega}{\partial T}\right|_V\bigg{/}\left.T\frac{\partial^2 \Omega}{\partial T^2}\right|_V.
\ea

\section{Numerical Result}
\begin{figure}[t]
\centering
\subfigure[]{\includegraphics[width=0.23\textwidth]{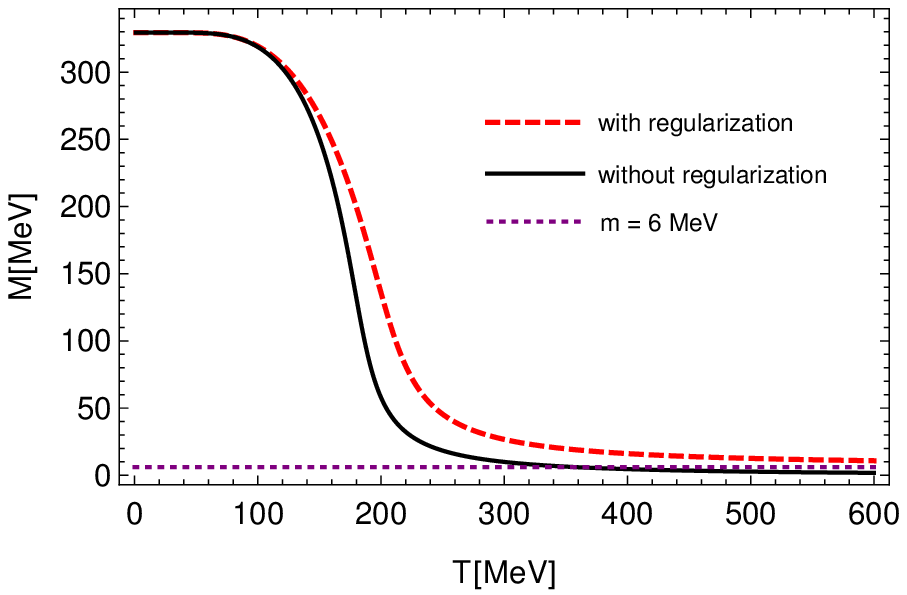}}
\subfigure[]{\includegraphics[width=0.23\textwidth]{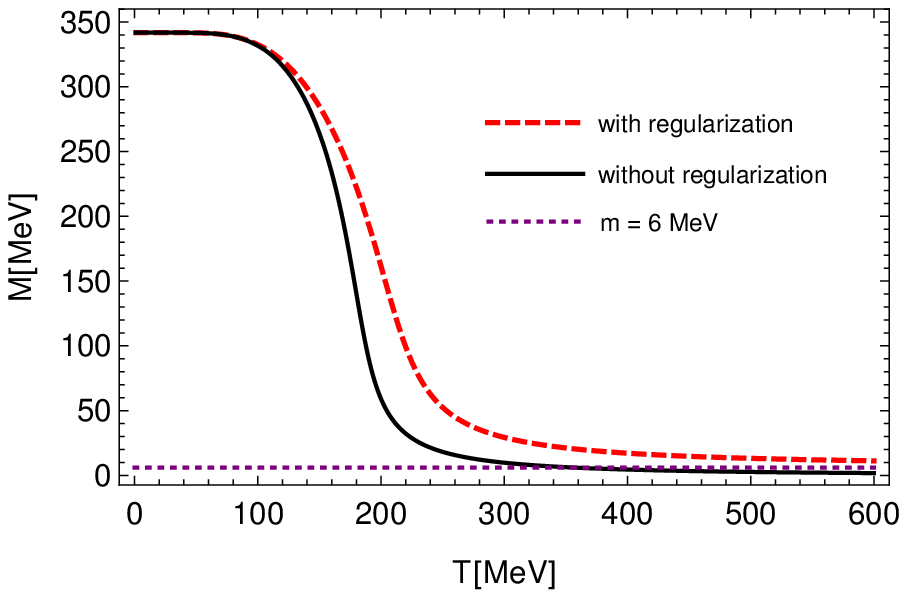}}
\subfigure[]{\includegraphics[width=0.23\textwidth]{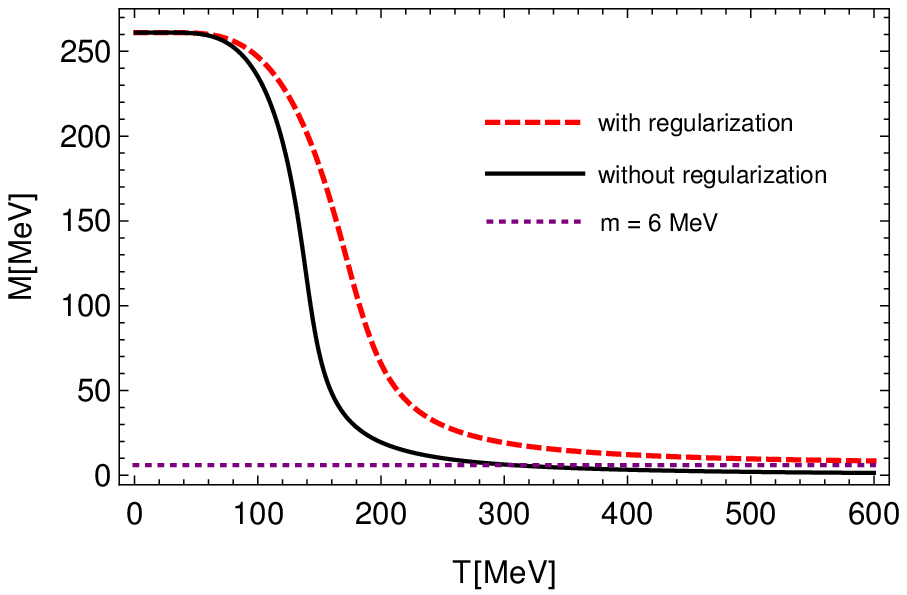}}
\caption{Constituent quark mass as a function of temperature with zero chemical potential. From panel (a)-(c), the three momentum hard cut off, three momentum soft cut off and Pauli-Villas regularization are used, respectively. In all the panels, the red dashed lines represent we are using the regularization on thermal part, the black solids line represent we are not using the regularization on the thermal part and the purple dotted lines are the current quark mass $m = 6$ MeV.}
\label{M}
\end{figure}
For simplicity, in the numerical calculations, we only consider the zero chemical potential case. It is obvious that the results should be equivalent to finite chemical potential cases. By solving the gap equation we calculated the constituent mass of quarks and several thermodynamical properties with zero chemical potential by using different regularizations. The constituent mass varies with temperature with different regularization schemes are showing in Fig.~\ref{M}. It is showing, by using all three different regularization schemes, the constituent quark mass is smaller than the current  quark mass at high temperature when we don't use the regularizations on the temperature part. Therefore, the chiral condensate is positive in this case, this is definitely an incorrect physics result.  When we use the regularizations on the temperature part, the chiral condensate goes towards zero when the temperature is increasing, this means the chiral symmetry is restoring at high temperature which is physically correct. By definition, the fact of constituent mass is always larger than the current mass in the framework of NJL model has been mentioned in Ref.~\cite{Buballa:2003qv}. In fact, this is not a model dependent result. In the chiral limit, the chiral symmetry is an exactly symmetry. The chiral condensate is negative in vacuum and increases to zero at some critical temperature.  Since the occurrence of chiral condensation is due to dynamical effect, similar feature should follow for nonzero current quark mass, that is, chiral condensate is always nonpositive at high temperature. Furthermore, the change of chiral condensate from negative value through zero to positive one would indicate that chiral symmetry is firstly restored and then broken again by only increasing temperature. This of course does not make sense. 

\begin{figure}[t]
\centering
\subfigure[]{\includegraphics[width=0.225\textwidth]{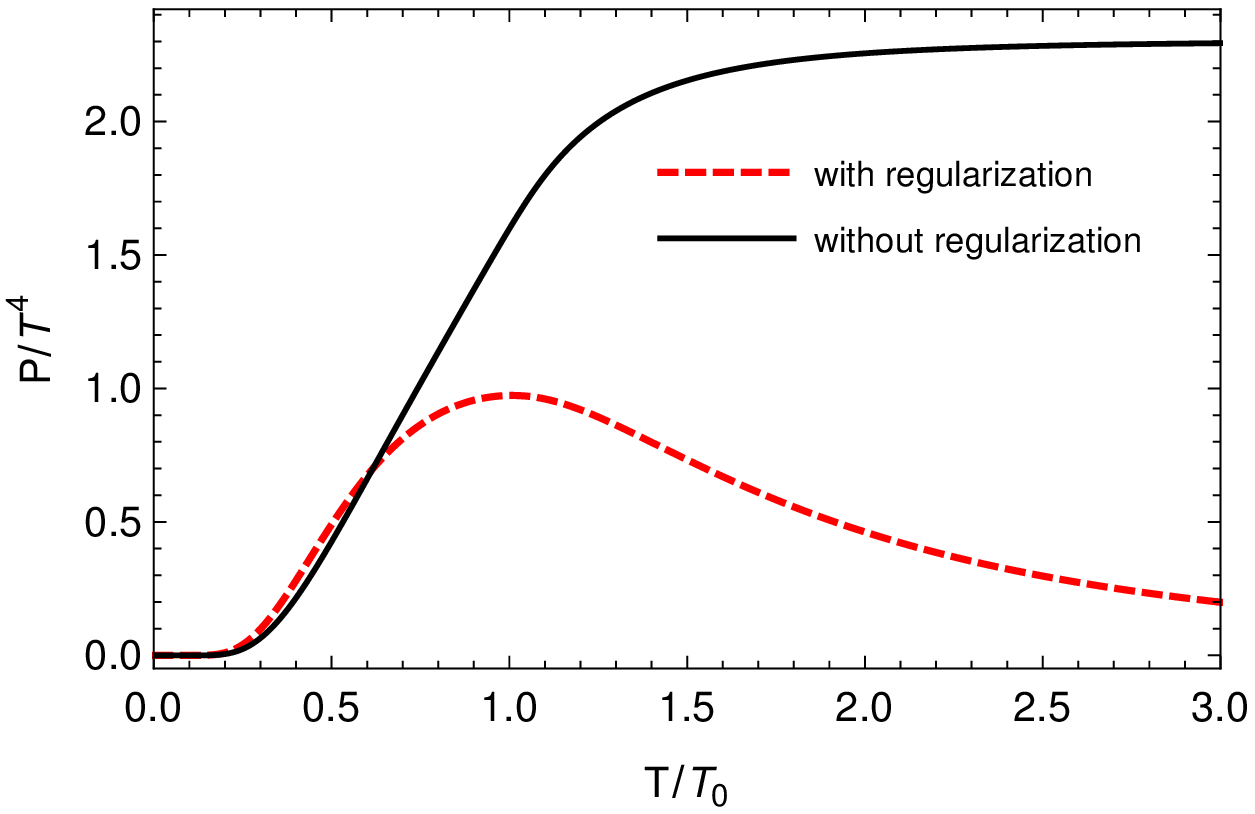}}
\hspace{0.01\textwidth}
\subfigure[]{\includegraphics[width=0.225\textwidth]{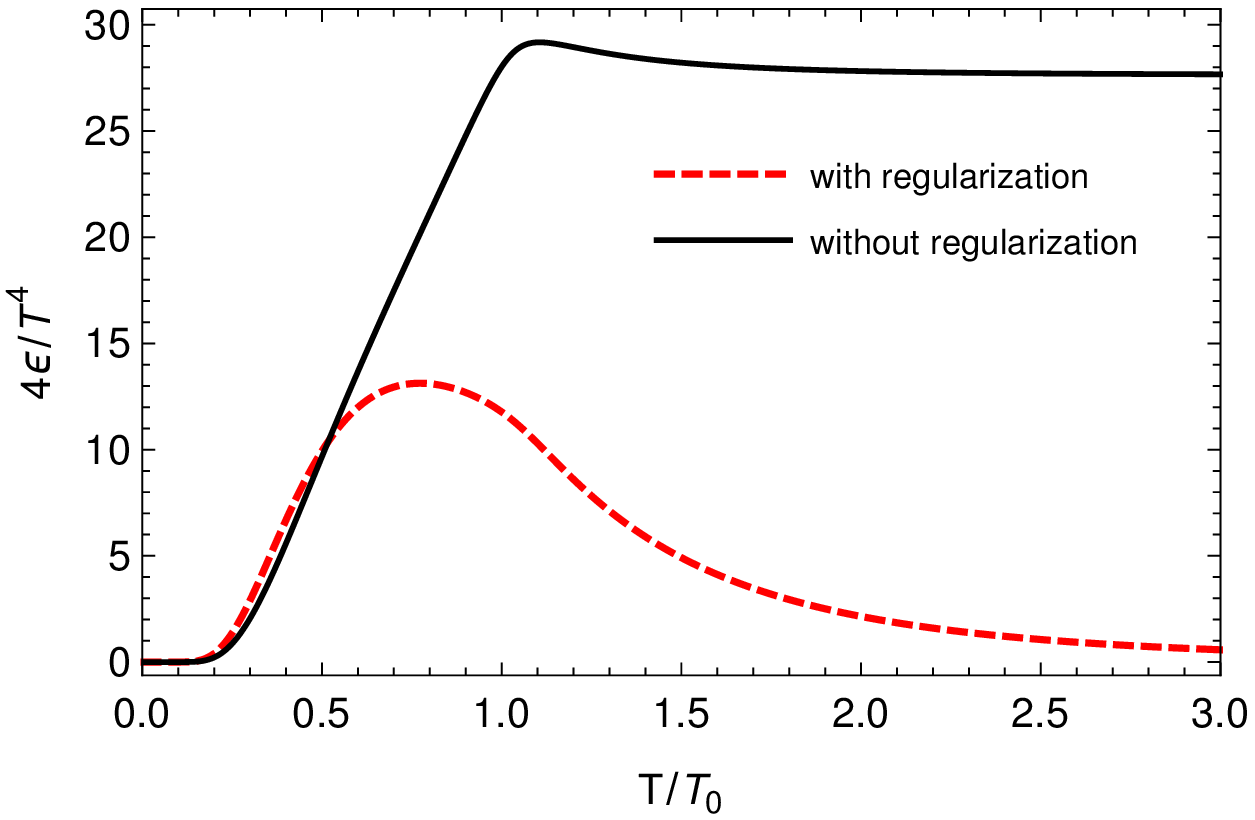}}
\subfigure[]{\includegraphics[width=0.225\textwidth]{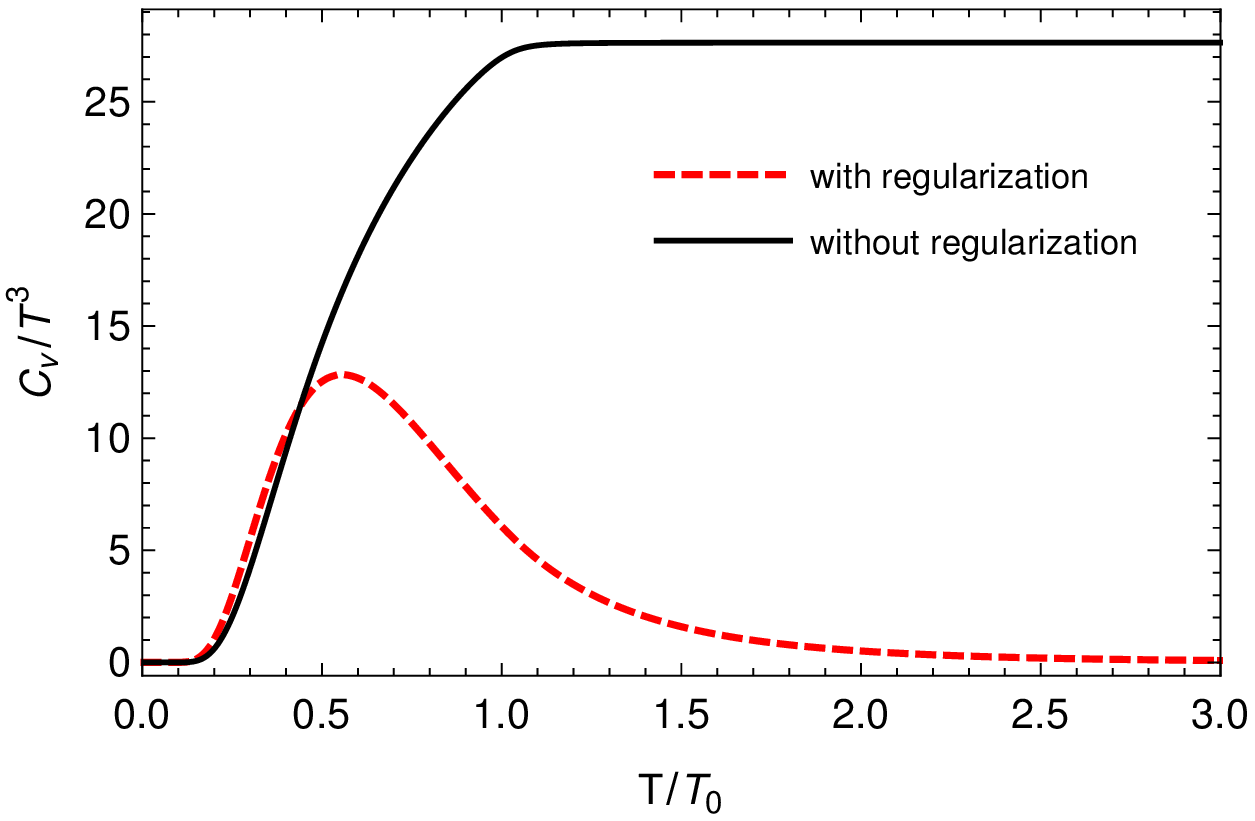}}
\hspace{0.01\textwidth}
\subfigure[]{\includegraphics[width=0.225\textwidth]{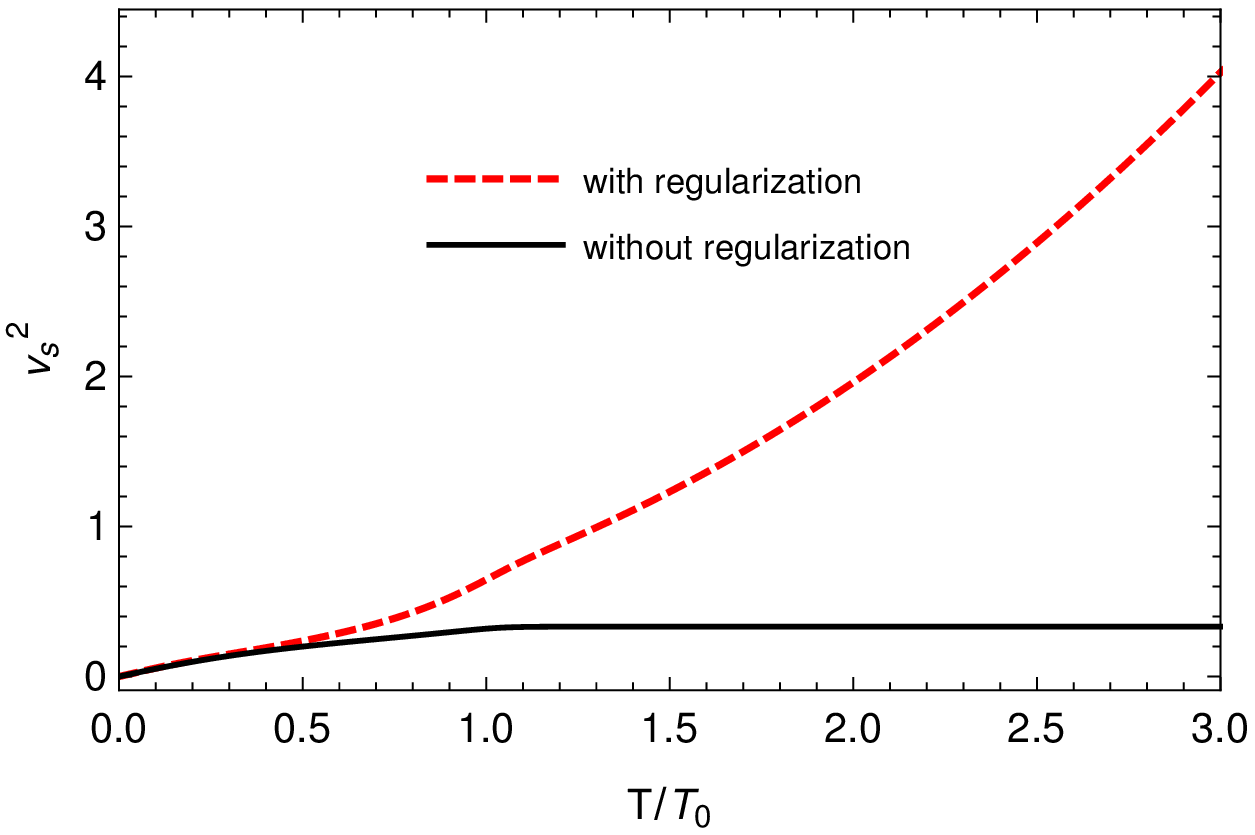}}
\caption{Dimensionless thermal properties as a function of normalized temperature with zero chemical potential by using three momentum hard cutoff. In all the panels, the red dashed lines represent we are using the regularization on thermal part, the black solid lines represent we are not using the regularization on the thermal part.}
\label{thermo-hard}
\end{figure}
\begin{figure}[t]
\centering
\subfigure[]{\includegraphics[width=0.225\textwidth]{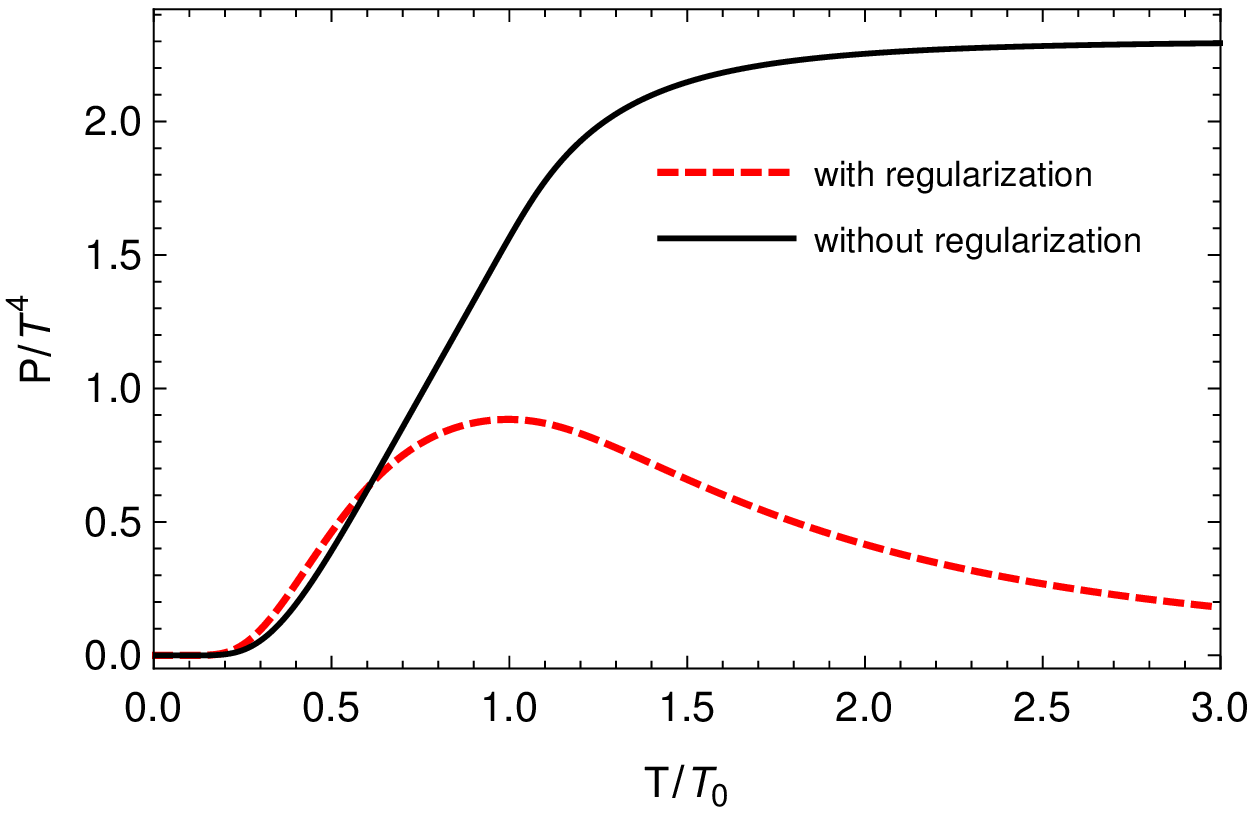}}
\hspace{0.01\textwidth}
\subfigure[]{\includegraphics[width=0.225\textwidth]{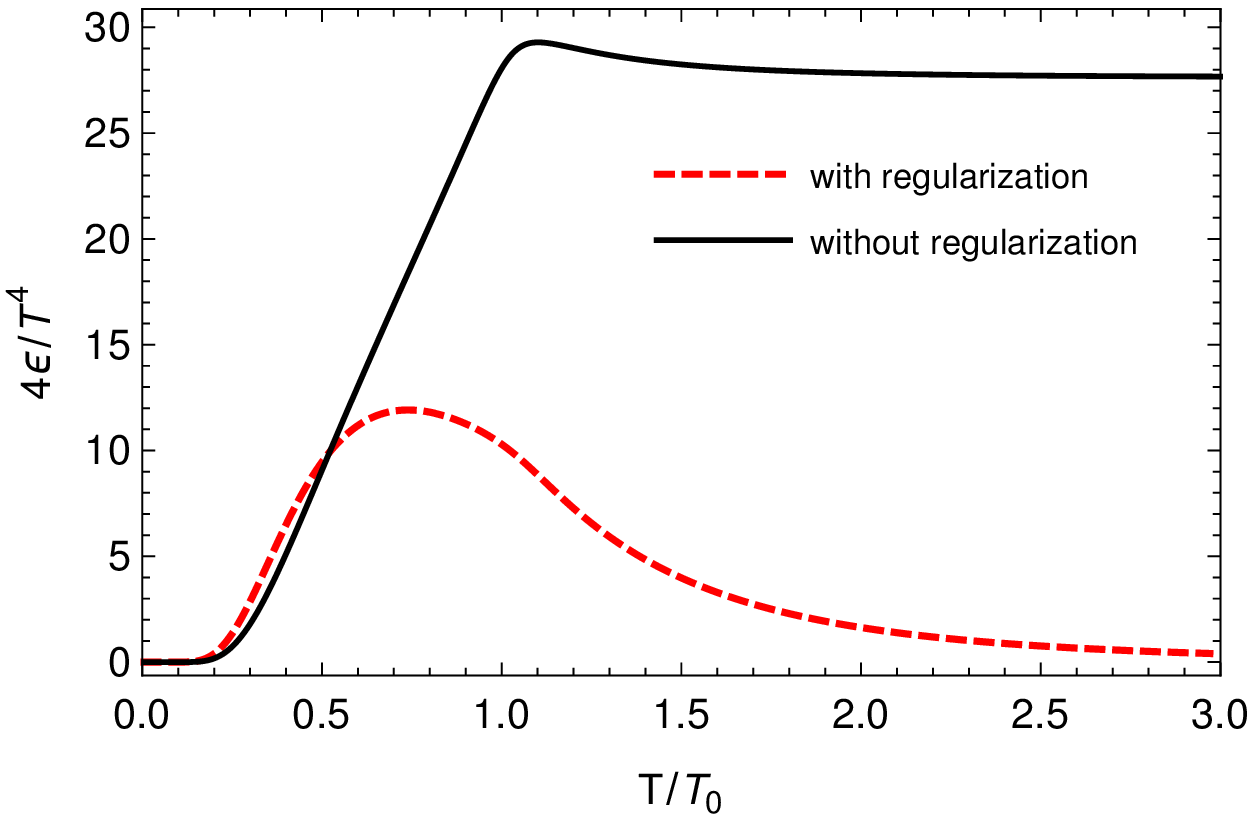}}
\subfigure[]{\includegraphics[width=0.225\textwidth]{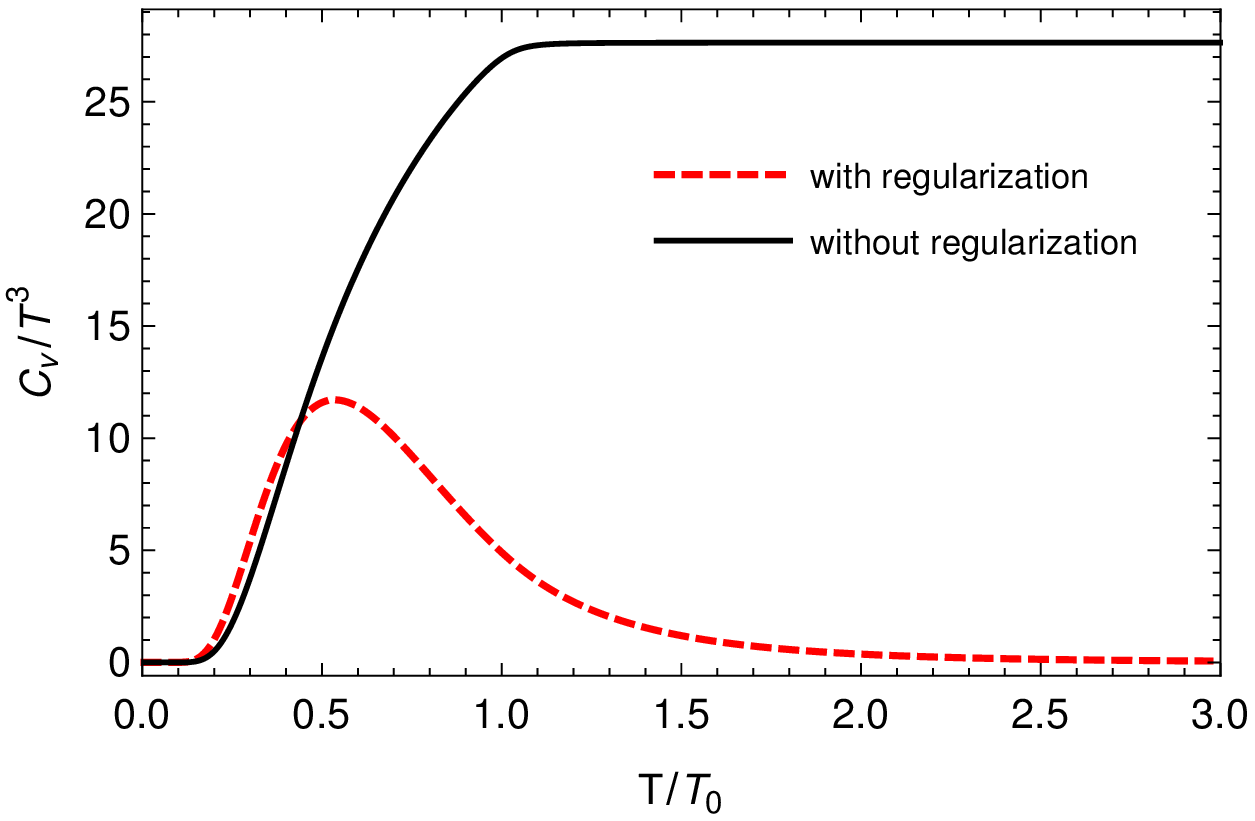}}
\hspace{0.01\textwidth}
\subfigure[]{\includegraphics[width=0.225\textwidth]{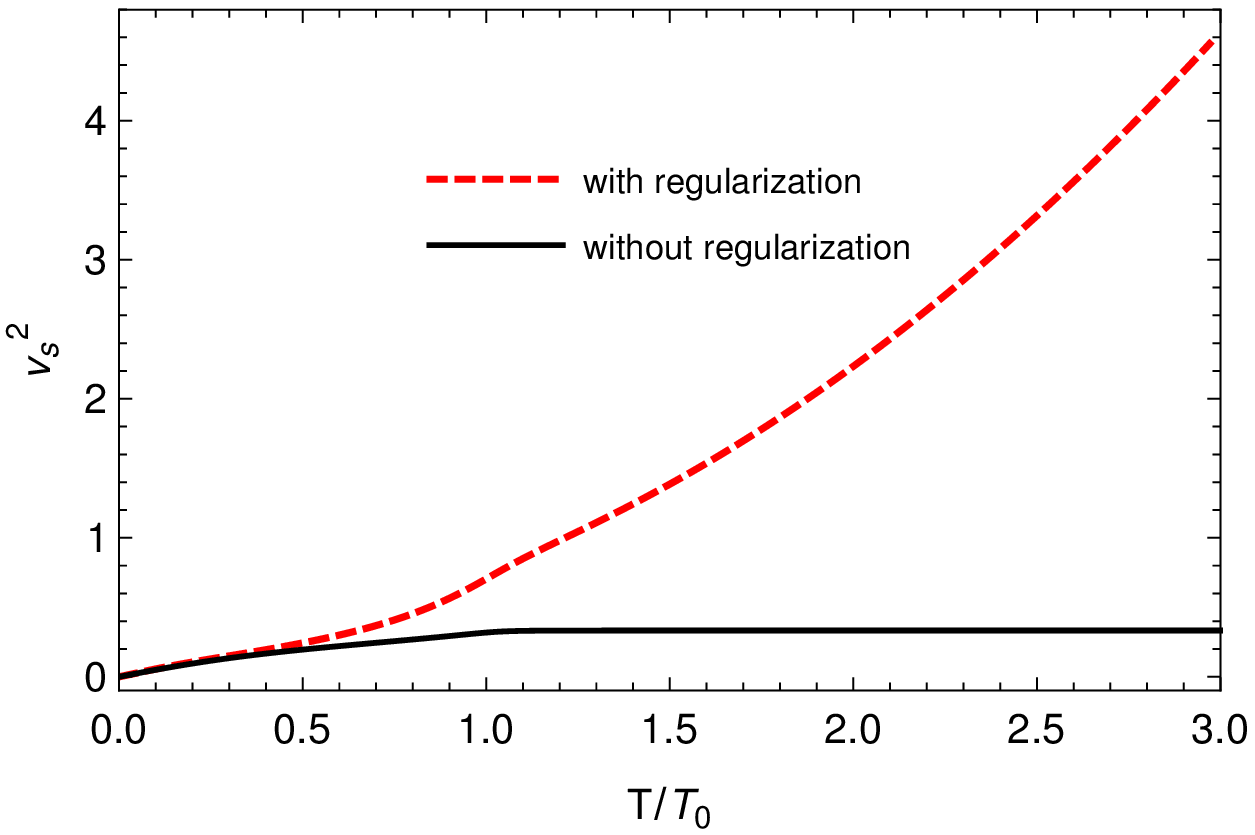}}
\caption{Dimensionless thermal properties as a function of normalized temperature with zero chemical potential by using three momentum soft cutoff. In all the panels, the red dashed lines represent we are using the regularization on thermal part, the black solid lines represent we are not using the regularization on the thermal part.}
\label{thermo-soft}
\end{figure}
\begin{figure}[t]
\centering
\subfigure[]{\includegraphics[width=0.225\textwidth]{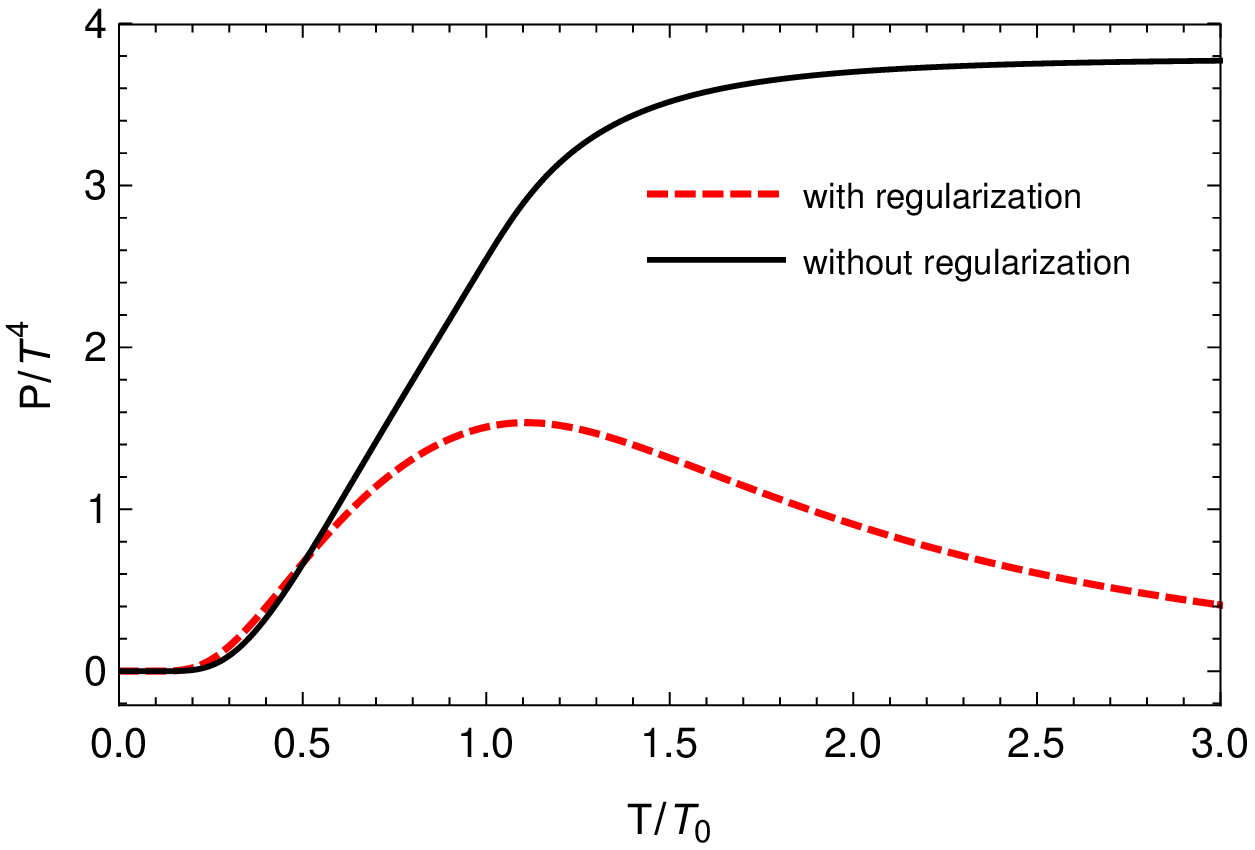}}
\hspace{0.01\textwidth}
\subfigure[]{\includegraphics[width=0.225\textwidth]{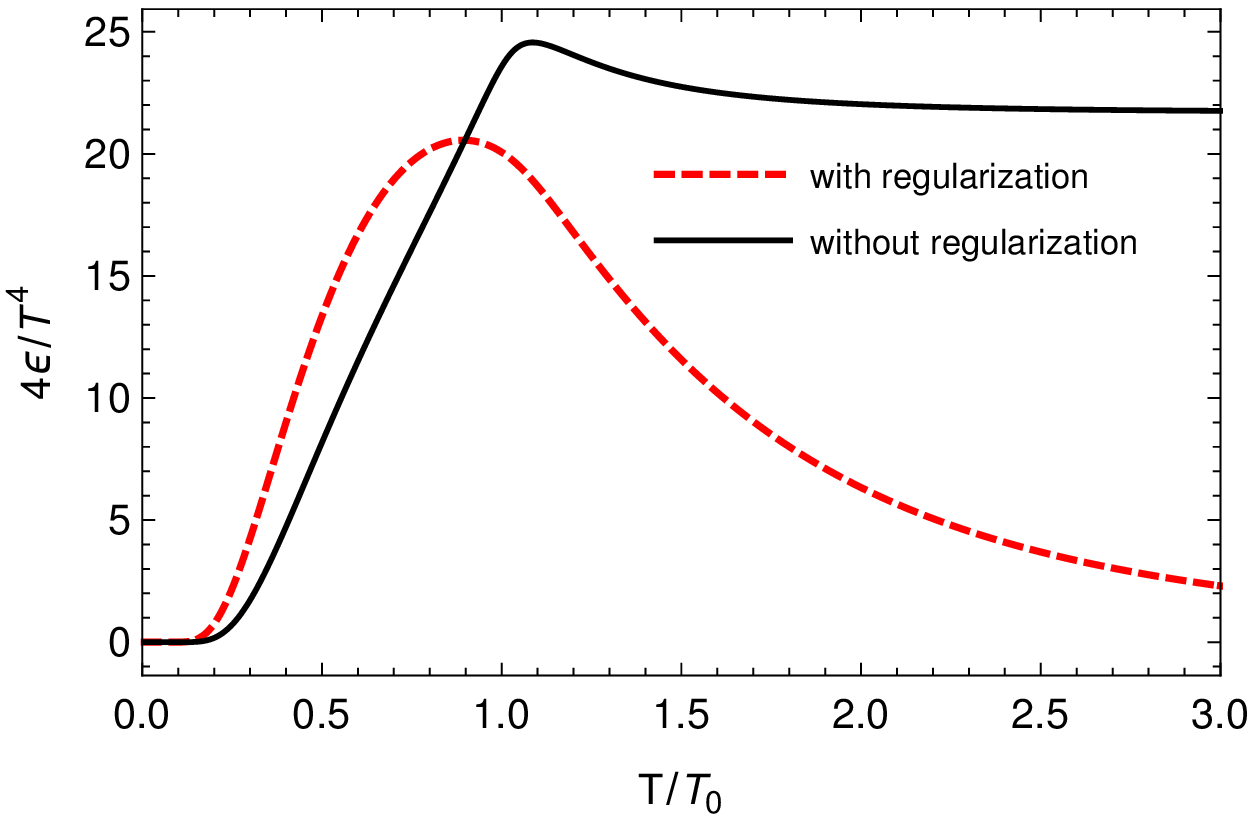}}
\subfigure[]{\includegraphics[width=0.225\textwidth]{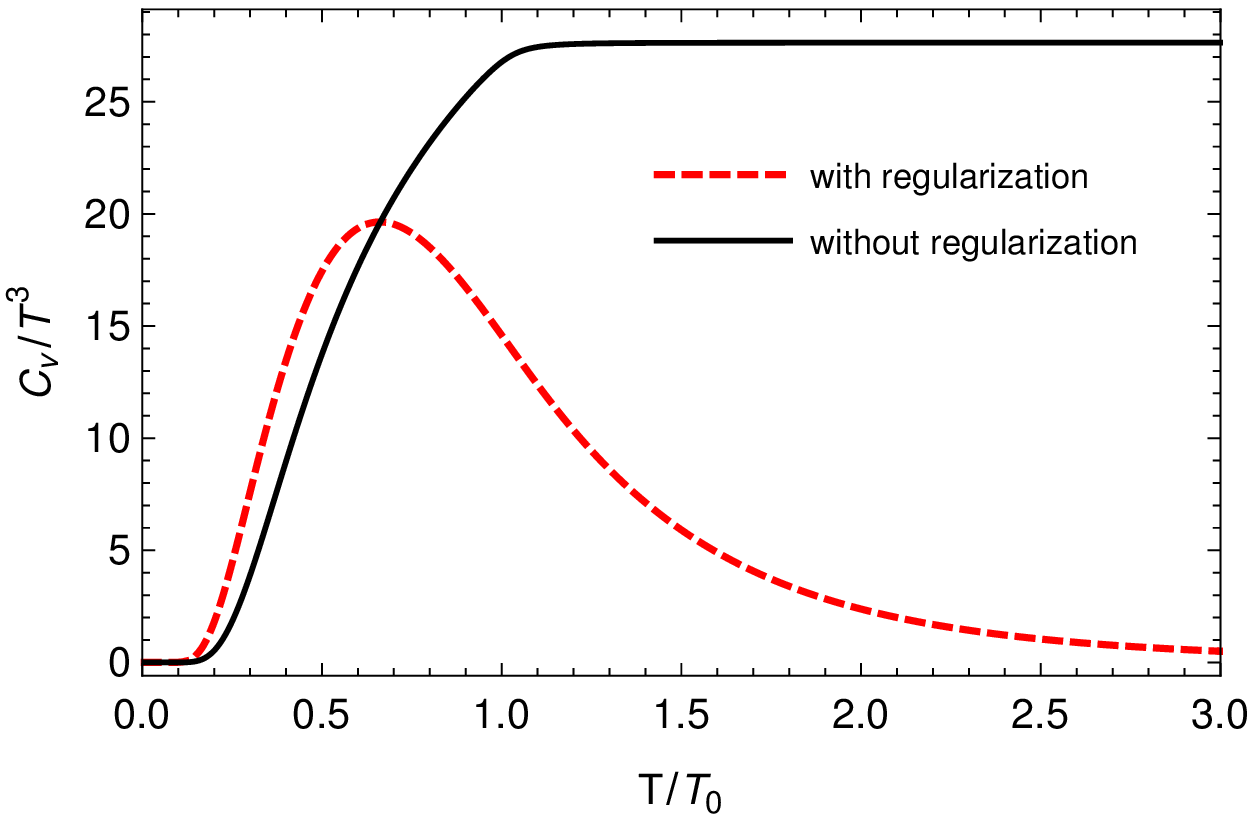}}
\hspace{0.01\textwidth}
\subfigure[]{\includegraphics[width=0.225\textwidth]{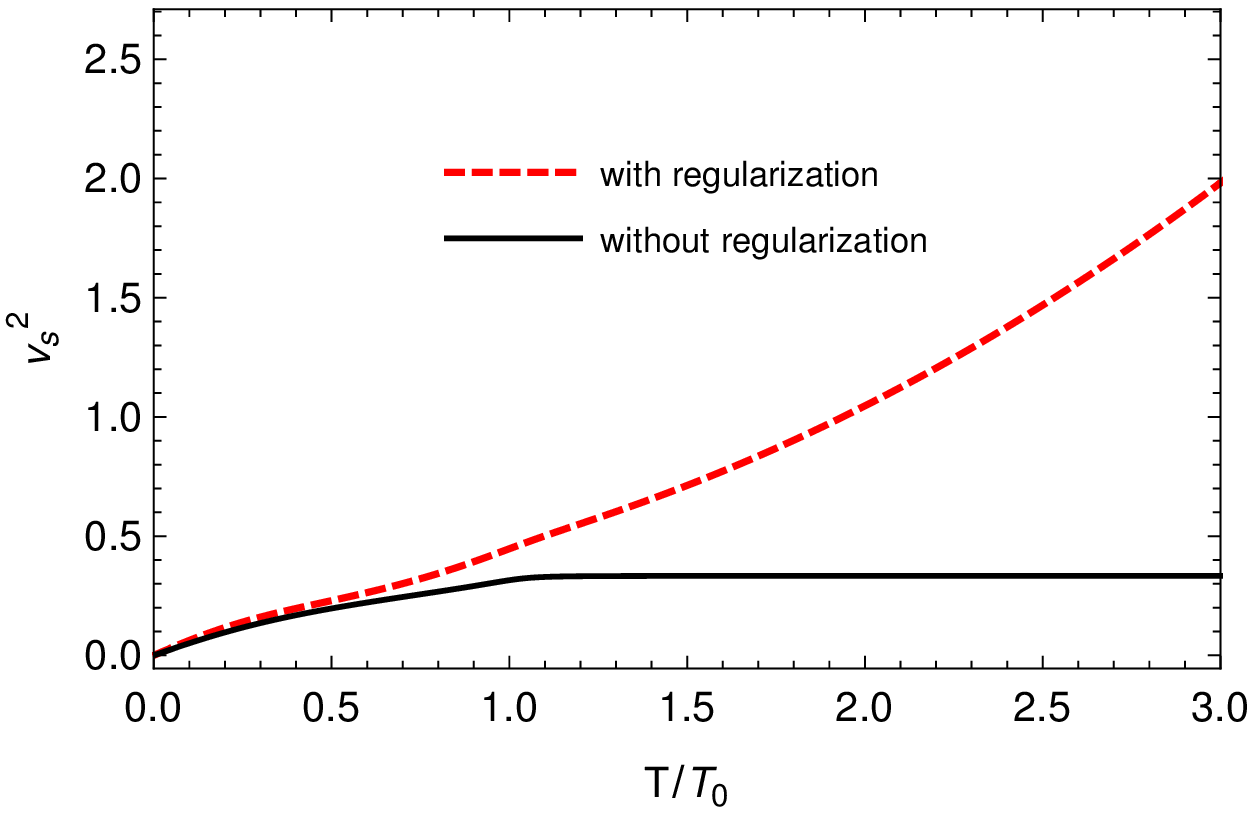}}
\caption{Dimensionless thermal properties as a function of normalized temperature with zero chemical potential by using Pauli-Villas regularization. In all the panels, the red dashed lines represent we are using the regularization on thermal part, the black solid lines represent we are not using the regularization on the thermal part.}
\label{thermo-PV}
\end{figure}
\begin{table}
\begin{tabular}{|c|c|c|}
\hline
~Regularization  Scheme~ & With(MeV) & Without(MeV) \\
\hline
three momentum hard cutoff&195&178  \\
\hline
three momentum soft cutoff&202&179   \\
\hline
Pauli-Villas regularization&172&138   \\
\hline
\end{tabular}
\caption{The value of $T_0$ from different regularization schemes. ``With"(``Without") is standing for using(not using) the regularization on the thermal part.}
\label{T0value}
\end{table}

In Figs.~\ref{thermo-hard}$-$\ref{thermo-PV}, we show the dimensionless quantities of the pressure, energy density, specific heat, and the square of sound velocity as a function of normalized temperature $T/T_0$ with different regularization schemes(hard cutoff, soft cutoff, Pauli-Villas regularization,  respectively) at zero chemical potential in the NJL model. $T_0$ is the critical temperature at zero chemical potential, the values by using different regularization schemes are given in Table~\ref{T0value}. By applying all three different regularization schemes, it is obvious to show the behavior with using and without using the regularizations on the thermal part are quite different for these quantities. It is well known when the temperature is increasing, the thermal quantities should be approaching the results of the free Fermi gas. The most trivial quantity is the speed of the sound, i.e., $v_s^2 = 1/3$ at high temperature. It is easy to see when we are not using the regularizations on the thermal part, the square of the speed of sound is approaching this value at high temperature, and it is incorrect when we are using the regularizations on the thermal part because it is larger than the speed of light when the temperature is high.  We have the same issue for other thermal quantities, no matter which regularization we are using. 

\section{Conclusion} In this paper, we discussed a fundamental question, whether we need to apply the regularization on the thermal part when we evaluate the physics quantities in the NJL model by considering two sets of physics qualities: the chiral condensate and thermodynamical quantities which is basically from the grand potential.  By selecting a wide range of temperature and three popular regularization methods (include both gauge covariant and gauge invariant schemes) in the NJL model, we find that to get a physical result, we need to use(not use) the regularization on the thermal part when we are evaluating the chiral condensate(grand potential) related physics quantities. A good example is the net baryon fluctuation in QCD matter which is very useful for studying the QCD phase transition, in the theoretical viewpoint, it purely depends on orders of derivative of the grand potential which we should not use the regularizations on the thermal part, e.g., in one of the author's previous papers~\cite{Li:2018ygx, Wang:2018sur}. On the other hand, for some quantities such as meson masses, which are directly related to the chiral condensate of quarks, we need to use the regularizations on the thermal part, e.g., Refs.~\cite{Liu:2018zag, Sheng:2020hge}. We could also come back to the question left in Refs.~\cite{Yu:2015hym} which we mentioned in the introduction section. For calculating the chiral condensate and related chiral phase transition temperature which are evaluated by the derivative of the chiral condensate, we need to use the regularization on the thermal part. Then the behavior of the critical temperature and chiral condensate will consistent with the results from lattice QCD~\cite{Braguta:2014ira} and Dyson-Schwinger Equations~\cite{Xu:2015vna}. A similar discussion about this specific problem is given in Ref.~\cite{Cui:2016zqp}.

Another thing we want to mention here is, although the numerical result is from the simplest two flavor NJL model, this should be correct for all NJL based models because of the model consistency. Also, we have covered the results for both gauge covariant and gauge invariant regularization schemes, our conclusion should be good for using all other regularization schemes. Therefore, the results should help understand the physics qualities of QCD matter, especially in the non-perturbation regime. Also, this work should be very helpful for beginners who are working on NJL based model projects. 
\acknowledgements
The work of X. Yu was supported by the Natural Science foundation of Jiangsu province under Grants No.BK20140523 and the start-up funding No.~15JDG042 of Jiangsu University. The work of X.W. was supported by the start-up funding No.~4111190010 of Jiangsu University and NSFC under Grant No.~11735007.

\end{document}